\documentclass[final]{svjour3}
\usepackage{graphicx}
\usepackage{rotating}
\usepackage{amssymb}
\usepackage{mathptmx}
\usepackage[numbers, sort&compress]{natbib}
\usepackage{bm}% bold math
\usepackage[mathlines]{lineno}% Enable numbering of text and display math
\usepackage{xcolor}

\newcommand{\eh}{$e^-h^+$}
\makeatletter
\journalname{Journal of Low Temperature Physics}
\bibpunct{[}{]}{,}{n}{}{,}
\begin{document}

\newcommand{\hdblarrow}{H\makebox[0.9ex][l]{$\downdownarrows$}-}
\title{Modeling of Impact Ionization and Charge Trapping in SuperCDMS HVeV Detectors}

\author{F.~Ponce$^{*1}$ \and W.~Page$^2$ \and P.L.~Brink$^3$ \and B.~Cabrera$^{1, 3}$ \and M.~Cherry$^3$ \and C.~Fink$^2$ \and N.~Kurinsky$^4$ \and R.~Partridge$^3$ \and M.~Pyle$^2$ \and B.~Sadoulet$^2$ \and B.~Serfass$^2$ \and C.~Stanford$^1$ \and S.~Watkins$^2$ \and S.~Yellin$^1$ \and B.A.~Young$^5$}

\institute{$^1$Department of Physics, Stanford University, Stanford, Ca 94305 USA\\
$^2$Department of Physics, University of California, Berkeley, Berkeley, Ca 94720 USA\\
$^3$SLAC National Accelerator Laboratory/KIPAC, Menlo Park, CA 94025 USA\\
$^4$Fermi National Accelerator Laboratory, Batavia, IL 60510, USA\\
$^5$Department of Physics, Santa Clara University, Santa Clara, CA 95053 USA\\
$^*$\email{ponce004@stanford.edu}\\
}
\maketitle

\begin{abstract}
A model for charge trapping and impact ionization, and an experiment to measure these parameters is presented for the SuperCDMS HVeV detector. A procedure to isolate and quantify the main sources of noise (bulk and surface charge leakage) in the measurements is also describe. This sets the stage to precisely measure the charge trapping and impact ionization probabilities in order to incorporate this model into future dark matter searches.

%A background model that incorporates bulk and surface charge leakage is presented for the SuperCDMS HVeV detector. We show that the time-varying optimal filter method distorts the Gaussian noise peak. We present a simplified model for impact ionization and charge trapping in these detectors and show how the model can be used to fit the fill-in regions between  observed quantization peaks corresponding to integer numbers of detected \eh~pairs.
%A background model that incorporates bulk and surface charge leakage is presented for the SuperCDMS HVeV detector. A simplified model for impact ionization and charge trapping is presented for these detectors and explain how the model can be used to fit the ``fill-in" regions between  observed quantization peaks corresponding to integer numbers of detected \eh~pairs.

\end{abstract}

\section{Introduction}
Non-baryonic matter, or Dark Matter (DM) comprises the vast majority of all matter in the universe. Many hypotheses exist speculating on the nature of DM with an increased interest in light DM (axion, dark photon, darksector)~\cite{essig13, alexander16, essig12, nelson11, holdom86} in recent years due to the lack of evidence for supersymmetry at the LHC. These new hypotheses have motivated R\&D efforts to develop high-resolution low-threshold detectors with single charge detection capabilities~\cite{romani18, tiffenberg17}. To achieve these requirements the SuperCDMS HVeV detector uses a bias voltage to convert a charge signal into an amplified phonon signal via the Neganov-Trofimov-Luke (NTL) effect~\cite{neganov81, luke88}. The ability to resolve single charges led to the observation of unwanted sub-gap IR (SGIR) photons, charge trapping and impact ionization~\cite{romani18} in our detectors. All of these potential noise sources require modeling to improve upon new DM constraints~\cite{agnese18} and/or interpret potential future discovery. In this paper we describe the capabilities of our system, present a model for charge trapping and impact ionization in the SuperCDMS HVeV detector, and outline an experiment to measure these parameters while accounting for the charge leakage background.

\section{Detector Operation and System Capabilities}
The SuperCDMS silicon HVeV detector is comprised of two concentric channels of several quasi-particle-trap-assisted electro-thermal-feedback transition-edge sensors (QETs) on one side and an Al parquet pattern on the other side of a $1{\times}1{\times}0.4$~cm$^3$ high-purity Si crystal (0.93~g)\cite{romani18}. Each QET sensor consists of four Al fins connected by a thin tungsten (W) film.  The total coverage (both channels) is 13\% of the top $1{\times}1$~cm$^2$ face. The Al parquet covers 20\% of the bottom $1{\times}1$~cm$^2$ face. During operation the QET sensors are DC biased with an Agilent 33210A Arbitrary Waveform Generator for stable operation within the W superconducting-to-normal transition and the Si crystal is DC biased with an iseg high voltage power supply (model NHR20 20R) between -160 and 160~V for NTL amplification.

The detector is mounted onto the mixing chamber stage of a Kelvinox-15 dilution refrigerator (DR) with a base temperature of $\sim$30~mK. The DR is fitted with a single mode fiber optic (FO) in order to illuminate the Al parquet side with 650~nm (1.91~eV) photons from a BNC Universal Pulse Generator (model 6040) laser. The laser is operated in pulse mode with an average number of photons $\lambda$ per pulse reaching the detector. The value of $\lambda$ depends on laser intensity, and the actual number of observed photons in an individual pulse follows Poisson statistics. The photon energy is sufficient to break \eh~pairs in silicon (band gap 1.14 eV) at a 1:1 ratio, which then undergo NTL amplification resulting in quantized peaks. 

For standard calibration acquisitions, a low background mode (limited to pulse pile-up) is achievable by triggering on the laser TTL. Further background mitigation is done to reduce SGIR through the FO by installing two KG-3 IR absorbing windows between the detector and FO output inside the DR. Additionally, background from leakage (dark) current is reduced to below 3 Hz by pre-biasing the Si crystal to +(-)160~V prior to acquisition at +(-)140~V~\cite{agnese18}. Thus an in-depth study of charge trapping and impact ionization can be carried out for the HVeV detector by using the laser as a signal source and operating in low background mode.

\section{Charge Trapping and Impact Ionization Model}
\begin{figure}[hbtp]
    \begin{center}
    \includegraphics[width=1\linewidth]{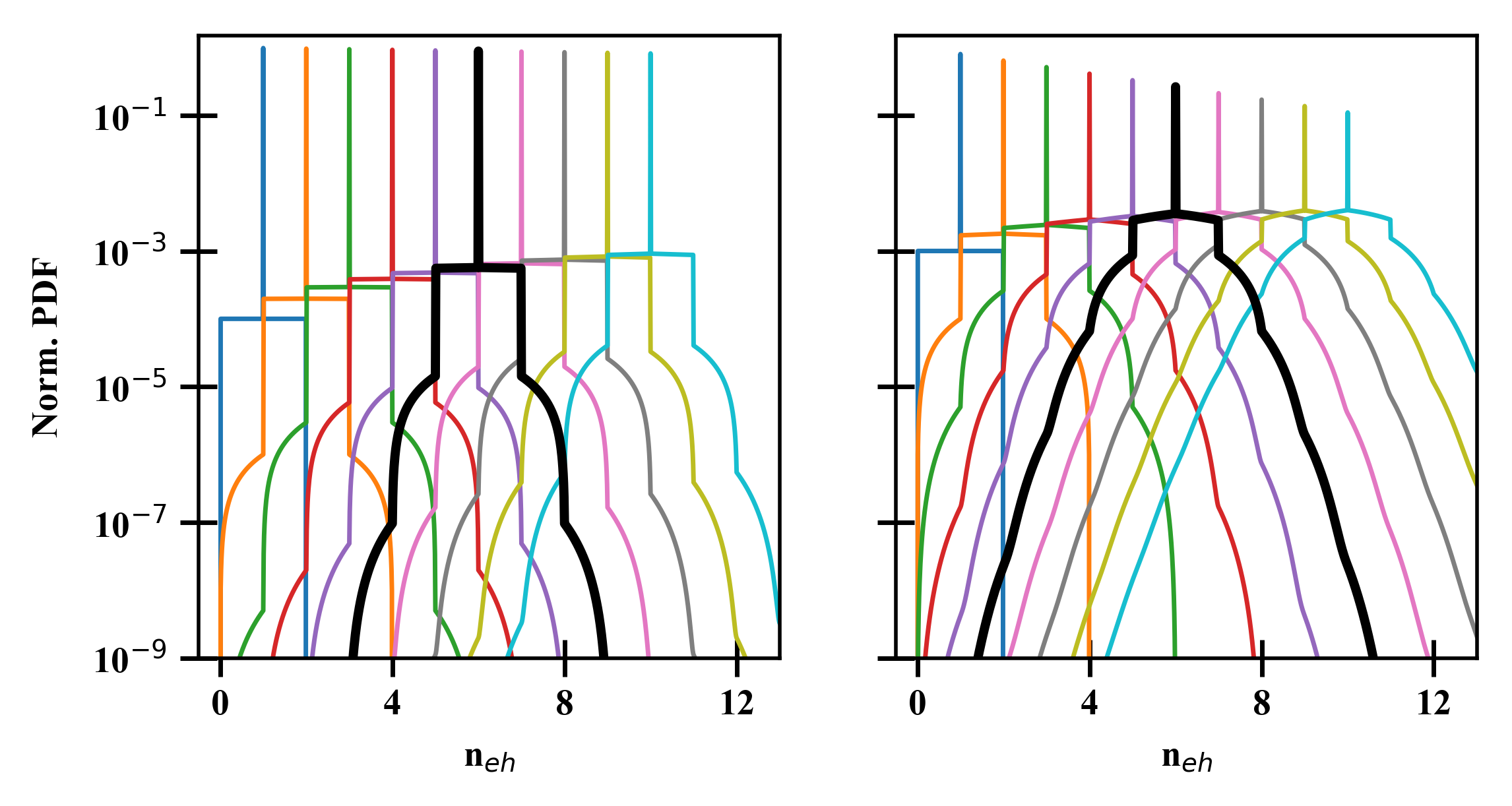}
    \caption{\footnotesize (color online) Normalized plot of $^{(m)}$h(x) for m $\leq$ 10 with an impact ionization and charge trapping probability both set to 1\% (left) and 10\% (right). The $^{(6)}$h(x) curve is plotted in black so that the higher order effects of Eq. (\ref{hpdf}) can be seen. For large values of m the amplitude of the main peak decrease.}
    \label{mhx}
    \end{center}
\end{figure}

The physical interaction of a single \eh~pair with the crystal is modeled as having some constant probability of charge trapping (effectively removing a charge), inducing impact ionization (effectively generating an additional charge), or having the original charges move through the crystal unhindered (resulting in a quantized signal). Each \eh~pair is assumed to undergo only one of the above interactions independent of all other \eh~pairs that were generated because of additional photons being absorbed. The corresponding probability distribution function (PDF) for a single \eh~pair is:
\begin{equation}
    ^{(1)}h(x) = A_-\Theta(x-0)\cdot\Theta(1-x)+ A_+\Theta(x-1)\cdot\Theta(2-x) + A_1\delta(x-1) \label{1hx}
\end{equation}
where A$_-$ is the charge trapping probability, A$_+$ is the impact ionization probability, A$_1$ = (1 $-$ A$_-$ $-$ A$_+$), $\Theta$ is the Heaviside function and x is the observed energy of the \eh~pair by the detector. The PDF for m \eh~pairs is calculated by convolving the single PDF with itself ``m" times using the recursive relationship:
\begin{equation}
    ^{(m)}h(x) = \int\limits_{-\infty}^{\infty}{}^{(1)}h(x')\cdot^{(m-1)}h(x-x')dx'
\end{equation}
Carrying out this recursive relationship gives the analytic solution for the m$^{th}$ peak PDF:
\begin{eqnarray}
    ^{(m)}h(x) &=& A_1^m\delta(x-m) +\nonumber\\
    & & mA_1^{m-1}A_-\Theta(x-m+1)\cdot\Theta(m-x) +\nonumber\\ 
    & & mA_1^{m-1}A_+\Theta(x-m)\cdot\Theta(m+1-x) +\nonumber\\ 
    & & \sum\limits_{i=0}^{m-2}\sum\limits_{j=0}^{m-i} \sum\limits_{n=1}^{m-i}A_{mijn}(n+m-j-x)^{m-i-j}\cdot\Theta(n+m-j-x)\cdot\Theta(x-m+j)
    \label{hpdf}
\end{eqnarray}
\begin{equation}
    A_{mijn} = \frac{A_1^iA_-^jA_+^{m-i-j}m!}{i!j!(m-i-j)!}\cdot\frac{(-1)^{m-i-n}(m-i)!}{n!(m-i-n)!}\cdot\frac{1}{(m-i-1)!} 
    \label{amijn}
\end{equation}
The first term in Eq. (\ref{hpdf}) represents all \eh~pairs traversing the Si crystal unhindered, the second (third) term represents a single \eh~pair undergoing charge trapping (impact ionization) and the fourth term represents all the other possible combinations of the three processes for all \eh~pairs.
The A$_{mijn}$ term is the combinatoric probabilities for charge trapping, and impact ionization given a fixed starting number of \eh~pairs (2 $\leq$ m). The first fraction comes from distributing (A$_1$ + A$_-$ + A$_+$)$^m$, the second fraction comes from integrating the $\Theta$ functions and the third fraction comes from the power rule for integrating polynomials of the function x$'$. Normalized plots of $^{(m)}$h(x) for m $\leq$ 10 and assuming an charge trapping/impact ionization probability of 1\% and 10\% is shown in Fig. \ref{mhx}. The effect of the 2$^{nd}$ - 4$^{th}$ terms in Eq. (\ref{hpdf}) is more pronounced for large m because of the increased probability that at least one interaction will occur. Additionally, we see that the 4$^{th}$ term in Eq. (\ref{hpdf}) cannot be ignored in the cases involving large impact ionization/charge trapping probabilities.

\begin{figure}[hbtp]
    \begin{center}
    \includegraphics[width=1\linewidth]{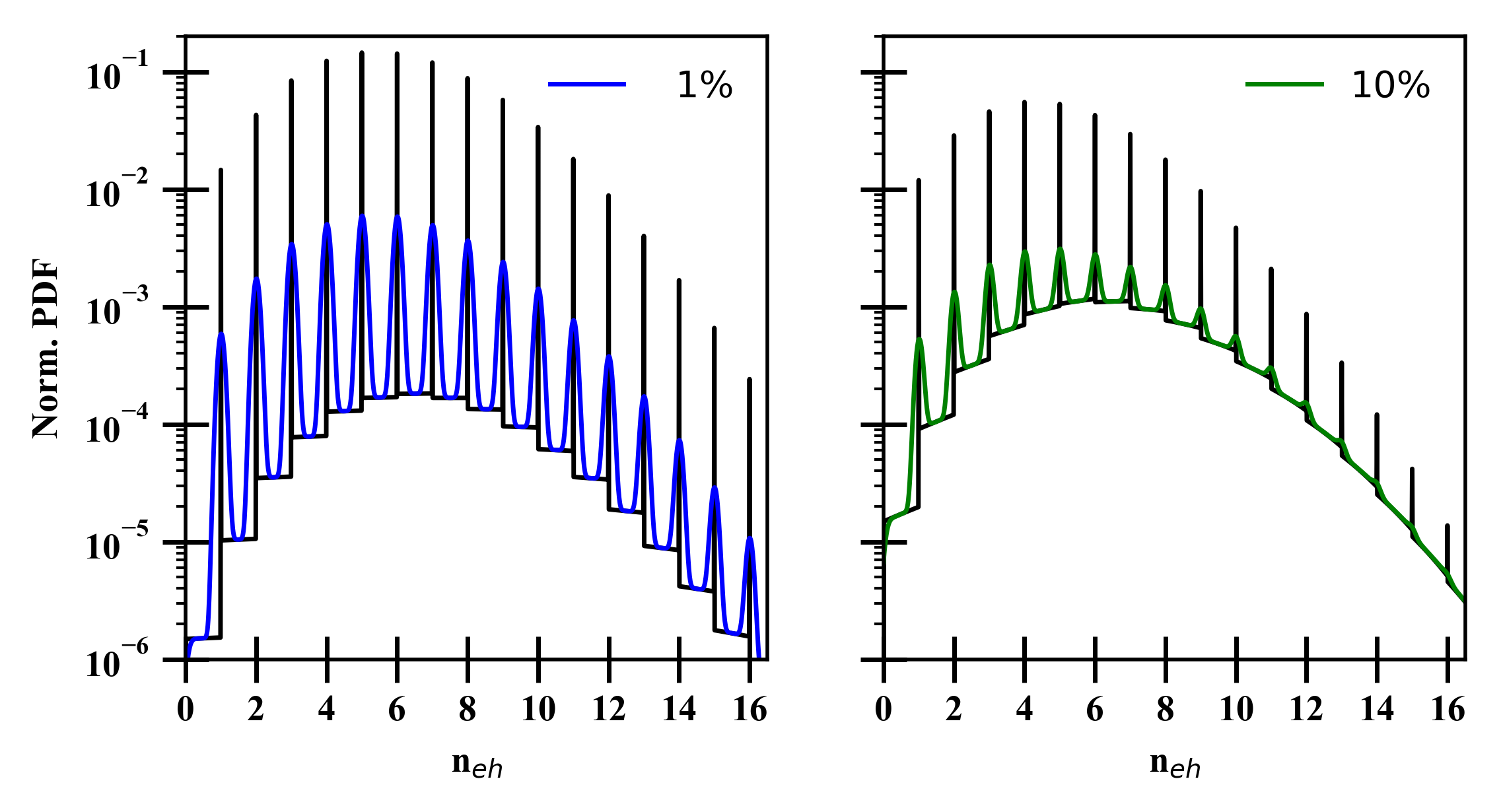}
    \caption{\footnotesize (color online) Expected detector response assuming an average of 6 \eh~pairs per pulse and a detector resolution of 0.1 \eh~pairs. The Gaussian detector response uses Eq. \ref{fullmodel} with an impact ionization/charge trapping probability both set to 1\% (left) and 10\% (right). %Excess events "fill-in" is expected between quantized peaks for large impact ionization/charge trapping probabilities.
    }
    \label{mhx_con}
    \end{center}
\end{figure}

To get the expected detector response to the pulsed laser (excluding the 0 \eh~pair peak), the PDF of Eq. (\ref{hpdf}) is numerically convolved with the Gaussian detector response for each m, scaled by the Poisson probability and summed together, 
\begin{equation}
    M(x) = \sum_{m = 1}^{m_{max}}P_m(\lambda)(^{(m)}h\circledast{}G(\sigma))(x) 
    \label{fullmodel}
\end{equation}
where $\lambda$ is the average number of photons per pulse. An example is shown in Fig. \ref{mhx_con} with a $\lambda$ of 6 \eh~pairs per pulse, and the same charge trapping/impact ionization probability from before of 1\% (blue), or 10\% (green). The 0 \eh~pair (noise peak) is excluded as it is a simple Gaussian peak with no charge trapping/impact ionization. Charge trapping and impact ionization is observed as ``fill-in" between the quantized peaks. The 1~\eh~pair peak provides the best signal to measure charge trapping and impact ionization because only one interaction is possible. In the case of impact ionization, the charge trapping from the 2~\eh~pair peak is the significant contaminant to the signal, and is mitigated by using a low laser intensity to suppress the 2~\eh~pair peak. In the case of charge trapping, the bulk/surface leakage in the noise peak is the only contaminate to the signal and is mitigated by using a modest laser intensity to maximize the signal source from the 1~\eh~pair peak, pre-biasing the Si crystal, and modeling the leakage background. 

\section{Leakage Background Model}
A detector's background can be classified into two categories: extrinsic (such as cosmic rays, environmental radioactivity and electromagnetic pick-up) and intrinsic (such as bulk impurities and surface contacts like the Al parquet). The effect of extrinsic sources can be mitigated by going deep underground and enforcing rigorous controls in the manufacturing and design of experiments. Since neither of these are options for our setup, a very stringent cut to the data can be used to remove high energy events by discarding entire sections of time. Thus extrinsic backgrounds are not considered in this model. The focus instead is on the intrinsic detector properties that result in single charge generation in the bulk and surface that undergo NTL amplification. 

\begin{figure}
    \centering
    \includegraphics[width=0.5\linewidth]{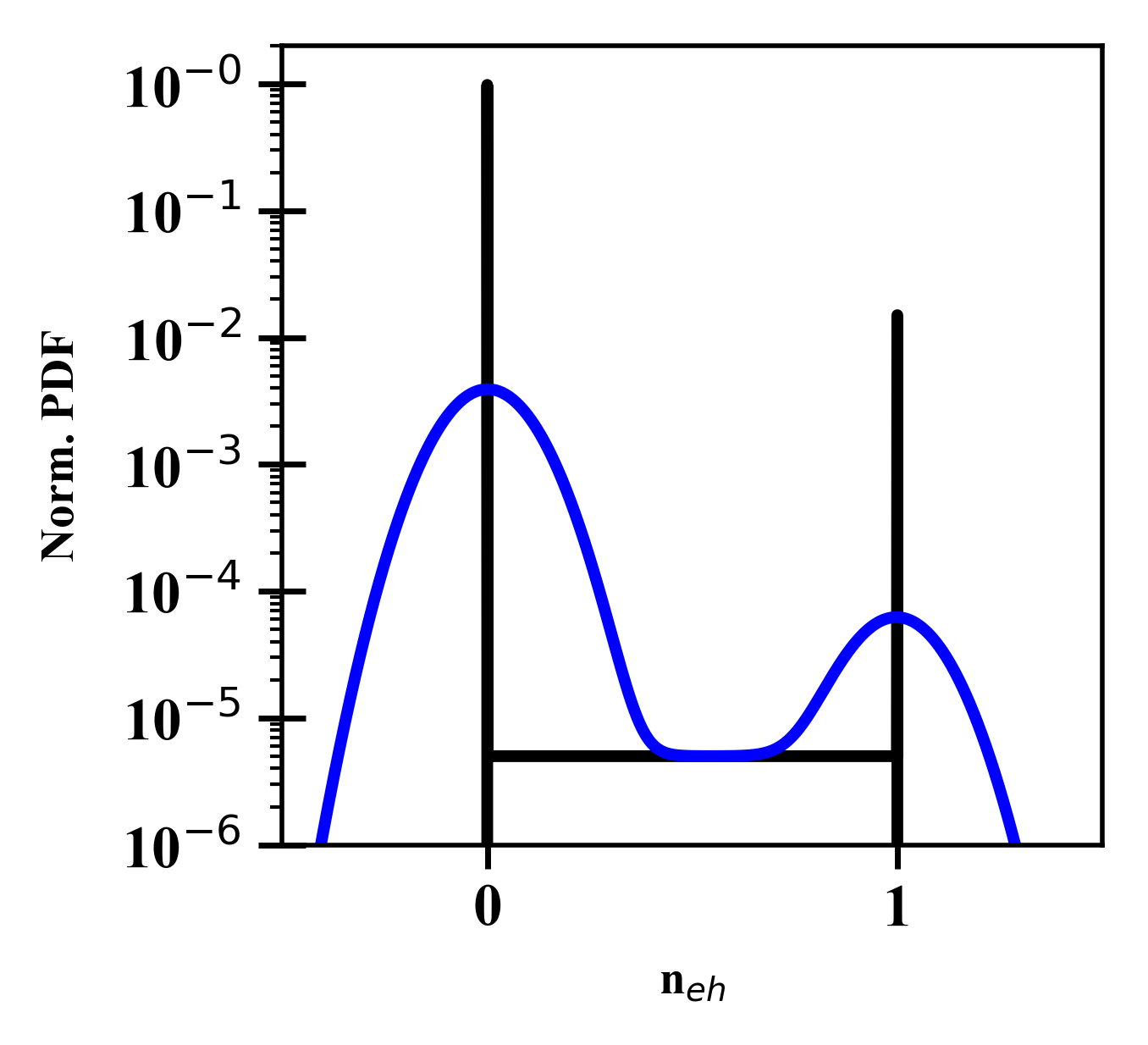}
    \caption{\footnotesize (color online) Plot of the leakage background PDF (black curve) convolved with the Gaussian detector response with $\sigma$ = 0.1 (blue curve).}
    \label{bkgd}
\end{figure}

Bulk leakage arises from impurities within the Si crystal that result in loosely bound charges at electrically neutral acceptor and donor atoms~\cite{burger84, fazzio85}, which can be excited by SGIR photons. Assuming that the excited impurities are uniformly distributed throughout the Si crystal the total NTL amplification of a freed single charge will depend on the starting position. This would be observed as a uniform distribution of energies between zero and full NTL amplification. Surface leakage is single charges excited from the crystal surface (i.e. fabrication contacts, defects) by any number of random processes and undergo full NTL amplification resulting in a quantized peak. The basic PDF model of the corresponding background leakage rate is,
\begin{equation}
    R(x) = R_{Surf}\delta(x - c_1) + \frac{R_{bulk}}{c_1 - c_0}\Theta(x - c_0)\cdot\Theta(c_1 - x)
    \label{bkgdrate}
\end{equation}
where R$_{Bulk}$ is the bulk leakage rate, R$_{Surf}$ is the surface leakage rate, c$_{0, 1}$ is the centroid of the quantized 0$^{th}$(1$^{st}$) \eh~pair peak, and x represents the observed energy of the detector. In the case of a low background mode laser acquisition the observed energy of leakage events depend on their relative arrival time (random) to the laser TTL. Such a measurement would not be accurate and thus the background model and acquisition method are at odds with one another.
%This model assumes a search for pulses so that the energies are well defined by the analysis routine and is thus not compatible with the low background mode. In low background mode, the observed leakage energies would depend on the relative arrival time of leakage events (random) to the laser TTL and introduce a systematic to the charge trapping/impact ionization probabilities.
%If there is no background then the relative arrival time of all non-zero \eh~pair laser events will cluster within $\pm\delta$t of 0~$\mu$s (TTL reference) while the zero \eh~pair events (noise) are distributed uniformly over the width of the search window~\cite{ponce19}. 

In order to bridge the acquisition method and background model a time-shifting optimal filter (OF)~\cite{golwala00, kurinski18} is implemented. The OF searches for (and takes) the maximum energy deposited over a fixed time window and marks the arrival time for each tagged event. Laser signals with a non-zero number of absorbed photons (non-zero \eh~pair events) cluster within $\pm\delta$t of 0~$\mu$s (TTL reference), while laser signals with zero absorbed photons are distributed uniformly over the width of the search window~\cite{ponce19}. The expected leakage background will be identifiable only when they arrive during the window in which no photons are absorbed (a zero \eh~pair event). The randomness of leakage events will place them outside the time range of the non-zero \eh~pair cluster. Thus the set of events outside $\pm\delta$t of 0~$\mu$s is properly represented by Eq.~\ref{bkgdrate} with two modifications. First a term that represents the number of events in which a leakage event is not observed (noise peak) is added and second the shape of this noise peak needs to account for the use of the OF~\cite{mancuso19}. To get the detector response for the leakage background Eq.~\ref{bkgdrate} is numerically convolved with the Gaussian detector response and the altered noise peak is added on,
\begin{eqnarray}
    B(x) &=& \frac{L_0Ne^{-\frac{x^2}{2\sigma^2}}}{\sqrt{2\pi\sigma^2}}\left(\frac{1}{2}\left(1 + erf\left(\frac{x}{\sqrt{2\sigma^2}}\right)\right)\right)^{N-1}\nonumber\\
    & & + \frac{L_{Surf}}{\sqrt{2\pi\sigma^2}}e^{-\frac{(x - c_1)^2}{2\sigma^2}}\nonumber\\
    & &+ \frac{L_{Bulk}}{2(c_1 - c_0)}\left(erf\left(\frac{x - c_0}{\sqrt{2\sigma^2}}\right) - erf\left(\frac{x - c_1}{\sqrt{2\sigma^2}}\right)\right)
    \label{bkgdpdf}
\end{eqnarray}
where the rates (R) are now probabilities (L) of a leak event arriving within the OF search window, N is a fit parameter representing the effective number of independent time bins at which noise can have its maximum upward fluctuation in the OF search window, $\sigma$ is the detector resolution, and L$_0$ = (1 - L$_{Bulk}$ - L$_{Surf}$) is the probability of no leakage events being observed. An example of the detector response for a leakage background assuming a detector resolution of $\sigma$ = 0.1, L$_{Bulk}$ = 0.5\%, L$_{Surf}$ = 1.5\% and N = 1 (Gaussian noise peak) is shown in Fig. \ref{bkgd} (blue curve). The flat bulk leakage background is observable between the noise and surface leakage peaks and represents the largest signal background to the measurement of charge trapping from the 1~\eh~pair laser peak. The PDF for this profile is shown in black with the noise peak taking the form of a delta function identical to the surface leakage. 

\section{Conclusion}
The characterization of charge trapping and impact ionization in the SuperCDMS HVeV detector is important to future DM search analyses. This model will enable these searches to consider the regions between quantized peaks and inform potential discovery or setting constraints. We outlined a charge trapping/impact ionization model for the SuperCDMS Si HVeV detector. The first \eh~pair peak is empirically determined to provide the best conditions to measure charge trapping/impact ionization probabilities. Additionally, a simplified model for the bulk/surface leakage background is described with a measurement procedure to reduce systematics.  

\begin{acknowledgements}
This work was supported in part by the U.S. Department of Energy and by the National Science Foundation. This document was prepared by using resources of the Fermi National Accelerator Laboratory (Fermilab), a U.S. Department of Energy, Office of Science, HEP User Facility. Fermilab is managed by Fermi Research Alliance, LLC (FRA), acting under Contract No. DE-AC02-07CH11359. SLAC is operated under Contract No. DEAC02-76SF00515 with the U.S. Department of Energy. The authors are also especially grateful to the staff of the Varian Machine Shop at Stanford University for their assistance in machining the parts used in this experiment.
\end{acknowledgements}

\end{document}